\documentclass[10pt,dvipsname]{iopjournal}

\usepackage{physics}
\usepackage[T1]{fontenc}
\usepackage{graphicx}
\usepackage{verbatim}
\usepackage[separate-uncertainty=true]{siunitx}
\usepackage{amsmath}
\usepackage{pifont} 
\usepackage{smartref}
\usepackage{subcaption}

\usepackage{hyperref}
\hypersetup{%
	colorlinks,
	citecolor=blue,
	linkcolor=blue,
	urlcolor=blue
}

\usepackage{tikz}

\newcommand{\Rb}{$^{87}$Rb}

\newcommand{\diff}{\mathrm{d}}
\newcommand{\e}{\text{e}}

\newcommand{\half}{\frac{1}{2}}

\DeclareRobustCommand{\cir}{\tikz[baseline=-0.6ex]\node[draw,scale=0.35,circle,very thick]{};}

\definecolor{bluemarkercolor}{rgb}{0.12156862745098039, 0.4666666666666667, 0.7058823529411765}
\definecolor{redmarkercolor}{rgb}{0.8392156862745098, 0.15294117647058825, 0.1568627450980392}

\begin{document}

\articletype{Paper} 

\title{Cloud parameter estimation for interacting BEC after time-of-flight}

\author{R. M. F.\,Andersen$^1$\footnote{Author to whom correspondence should be addressed.}, S.\,Frederiksen$^1$, L.\,Stokholm$^1$, I.\,Zebergs$^1$, M.\,Kristensen$^1$, C.\,A.\,Weidner$^2$, J.\,J.\,Arlt$^1$}

\affil{$^1$ Center for Complex Quantum Systems, Department of Physics and Astronomy, Aarhus University, Ny Munkegade 120, DK-8000 Aarhus C, Denmark.}
\affil{$^2$ Quantum Engineering Technology Labs, H. H. Wills Physics Laboratory and School of Electrical, Electronic, and Mechanical Engineering, University of Bristol, BS8 1FD, United Kingdom.}

\email{rmfa@phys.au.dk}

\begin{abstract}

Experiments on Bose-Einstein condensates at finite temperature typically extract the system parameters, such as temperature, atom number, and condensed fraction from time-of-flight images taken after a free expansion time. 
This paper systematically examines the effect of repulsive interactions between the condensed and thermal atoms in partially condensed clouds on the expansion profile of the thermal cloud. An analytical expression for the expansion can be obtained only if the interactions between the Bose-Einstein condensate and thermal atoms are neglected, resulting in a Bose-enhanced distribution for the thermal component. Here, the deformation of the cloud due to interactions and the effects on estimated parameters are investigated by simulating the expansion using a ballistic approximation. By fitting the simulated expansion profiles with a Bose-enhanced distribution, the errors of using such a fit are estimated, and the results are explained phenomenologically. The simulation was also used as a fitting function for experimental data, showing better agreement of the extracted condensed fraction with the semi-ideal model than results from a Bose-enhanced fit. 
\end{abstract}

\

\section{Introduction}
\label{sec:Introduction}

Absorption imaging has traditionally been one of the most important detection techniques in ultracold atomic physics and has played a central role in the observation and characterization of Bose–Einstein condensation~\cite{img6}. Despite its widespread use, the accurate determination of system parameters remains a nontrivial challenge, with systematic uncertainties arising from both experimental and theoretical considerations~\cite{img1, img3, img4}. Over the years, several variants of absorption imaging have been developed to mitigate these limitations~\cite{img2, img4, img5, img7, img8, img9}. In addition, partially destructive yet resonant techniques such as partial-transfer absorption imaging have been introduced, enabling repeated measurements of atomic clouds with reduced perturbation~\cite{Campbell2012}.

Generally, in absorption imaging, resonant light incident on the atomic cloud is absorbed, casting a shadow that can then be imaged on a camera. When taking \emph{in situ} absorption images of trapped clouds of ultracold atoms, however, the optical densities are often too high for any meaningful amount of light to penetrate the cloud. Additionally, the small spatial extent of in-situ clouds requires special imaging optics for sufficient resolution. Consequently, the standard procedure is to release the cloud in free fall for a short time of flight prior to imaging. This results in an expansion of the cloud and thus a reduction in the optical density, allowing for more precise measurements. For a detailed discussion of absorption imaging analysis and noise estimation, see Ref.~\cite{img1, img3, img4}.

To extract precise information from absorption images, e.g., atom number and temperature, one typically needs a model for the cloud expansion in time-of-flight. This is a relatively simple problem for ideal gas clouds released from a harmonic trap~\cite{thermalTOF, Vibel}, which can be solved analytically and provides a good approximation in many cold atom experiments since the densities of the clouds are low enough for interactions to be negligible. However, when a bosonic gas approaches degeneracy and obtains a non-negligible Bose-Einstein condensate (BEC) component, a macroscopic number of atoms populate the ground state, resulting in a sharp increase in density. As a result, the ideal gas assumption breaks down, and atom-atom interactions must be taken into account.

When one considers a cloud consisting of both Bose-condensed and thermal atoms, the thermal component can be approximated as an ideal gas, and often interactions between the two components are ignored, thus decoupling the BEC and the thermal atoms. However, the high-density condensed atoms, which break the ideal gas assumption, interact with all atoms, including the thermal atoms. This leads to a more complicated system where not even an approximate solution to free expansion is available analytically. In this paper, a numerical model is employed to estimate the effect of these interactions on the thermal cloud's expansion, specifically considering two key parameters, temperature ($T$) and thermal atom number ($N_{th}$)~\cite{MalthePhD}.

The paper is structured as follows. Section~\ref{sec:ExpandingClouds} reviews the expansion of pure thermal clouds and pure BECs. The following section, Sec.~\ref{sec:ExpandingMultimodalCloud} discusses the simulation of the expansion of partially condensed clouds, including the in-trap forces that arise when the condensed atoms interact with other atoms in the cloud. Section~\ref{sec:Errors} discusses the errors that arise in typical absorption imaging when these interactions are not taken into account, and Section~\ref{sec:Fits} shows that the model can be used to fit experimental data and extract more accurate temperatures and atom numbers. Section \ref{sec:Conclusion} concludes the paper.

\section{Expanding pure thermal clouds and Bose-Einstein condensates}
\label{sec:ExpandingClouds}

This section provides a brief review of the free expansion of pure thermal clouds and pure BECs, as well as the fitting technique used to extract information from cloud images. It serves as an introduction to the necessary methods to model partially condensed clouds. 

\subsection{Freely Expanding Pure Thermal Cloud}
\label{subsec:Freely Expanding Thermal Cloud}

Non-interacting bosonic particles with mass $m$ at temperature $T$ and chemical potential $\mu$ are described according to the Bose-Einstein distribution~\cite{Schroeder2014}
\begin{align}\label{eq. boseeinsteindistribution}
 n_{th}(\textbf{r}, \textbf{p})=\frac{1}{\e^{(\textbf{p}^2/2m+V(\textbf{r})-\mu)/k_BT}-1},
\end{align}
where $n_{th}(\textbf{r}, \textbf{p})$ is the average phase-space density of particles at position $\textbf{r}$ and with momentum $\textbf{p}$ in a trap potential $V(\textbf{r})$. From now on, it is assumed that $V(\textbf{r})$ is harmonic with trap frequencies $\omega_{i}$.

The spatial distribution $n_{th}$ can be found by integrating over the momentum-space with volume element $\diff^3 \textbf{p}/(2\pi\hbar)^3$. Here, the interactions between atoms are neglected, and after being released from the trap, the expansion is ballistic. To find the density of atoms $n_{th}(\textbf{r}, t)$, the probability of atoms arriving at $\textbf{r}$ from all parts of the in-trap cloud after a time of flight $t$ must be summed. This is accomplished by integrating over all momenta and calculating the initial position of an atom with each given momentum $\textbf{p}$ that ends up at $\textbf{r}$, which is $\textbf{r} - \textbf{p}t/m$, giving
\begin{align}\label{eq. time dependent thermal cloud}
    \nonumber n_{th}(\textbf{r},t) &= \int\frac{\diff^3 \textbf{p}}{(2\pi\hbar)^3}\frac{1}{\e^{(\textbf{p}^2/2m + V(\textbf{r}-\textbf{p}/m\cdot t) - \mu)/k_BT}-1}\\
    &=\left(\frac{mk_BT}{2\pi\hbar^2}\right)^{3/2}g_{\frac{3}{2}}\left(\Tilde{z}\e^{ -\sum_{i=1}^3 x_i^2/2w_i^2} \right),
\end{align}
where $g_{\frac{3}{2}}$ is the polylogarithm function defined by
\begin{align}
    g_\gamma(x) = \sum_{i=1}^{\infty}\frac{x^i}{i^\gamma},
\end{align}
and $w_i(t) = \sigma_i\sqrt{1+\omega_i^2t^2}$ are the widths after time-of-flight with in-trap widths $\sigma_i = \sqrt{\frac{k_BT}{m\omega_i^2}}$ and fugacity $\Tilde{z}=\exp(\mu/k_BT)$. 

Equation~\eqref{eq. time dependent thermal cloud} describes a Bose-enhanced Gaussian distribution, which is characterized by a sharp central peak for low temperatures. At high temperatures, only the first order of the polylogarithm function expansion contributes, and the distribution approaches a standard Gaussian distribution.

\subsection{Freely Expanding Pure BEC}
\label{subsec:Freely Expanding BEC}

The condensed part of the cloud is typically well described by the Gross-Pitaevskii equation~\cite{Pethick2008, Gross1961, Pitaevskii1961}. The Thomas-Fermi approximation neglects the kinetic energy in the Gross-Pitaevskii equation, and provides a good approximation for the BEC density distribution in-trap. The regions where the approximation fails coincide with the lowest condensed atom densities (such as near $T_c$ and on the cloud edge).

In this approximation, the BEC density $n_0$ of a condensed cloud with $N_0$ atoms is given by~\cite{Pethick2008} 
\begin{align}\label{eq. Free BEC}
    n_0(\textbf{r}) = \frac{15}{8\pi}\frac{N_0}{R_1R_2R_3}\cdot \Theta \left(1 - \sum_{i=1}^3\frac{\textbf{r}_i^2}{R_i^2}\right),
\end{align}
where $\Theta$ is the Heaviside step function, $R_i = \sqrt{2\mu/m\omega_i^2}$ are the Thomas-Fermi radii, and $\mu$ is the chemical potential, which is given in the Thomas-Fermi approximation by $\mu = \half \left(15N_0a/a_{\text{ho}}\right)^{2/5}\hbar \Bar{\omega}$, where $a$ is the $s$-wave scattering length, $a_{\text{ho}}=\sqrt{\frac{\hbar}{m\Bar{\omega}}}$ is the characteristic harmonic oscillator length, and $\Bar{\omega}=(\omega_1\omega_2\omega_3)^{\frac{1}{3}}$ is the geometric mean of the trap frequencies.

During time-of-flight, such a BEC cloud retains the same profile but with time-dependent radii~\cite{CastinDum, expandingBEC2} $R_i(t)=\lambda_i(t)\cdot R_i$. The scaling parameters $\lambda_i(t)$ with respect to  the initial radii can be found by solving the coupled differential equations
\begin{align}
    \Ddot{\lambda_i}=\frac{\omega_i^2}{\lambda_1\lambda_2\lambda_3\cdot \lambda_i}.
\end{align}
This provides the expansion of a pure BEC, which is easily obtained numerically. The equation predicts a reversal of the aspect ratio of the BEC, since the expansion rate along a particular axis is inversely proportional to its size along that axis~\cite{img6}. 

\subsection{Analytic evaluation of partially condensed Clouds}
\label{subsec:ColumnDensities}

Absorption imaging yields only the two-dimensional column densities $\Tilde{n}(x_1,x_2,t)=\int n(\textbf{r},t)\diff x_3$ of clouds rather than the full three-dimensional distributions. For BEC, the column density from integrating Eq.~\eqref{eq. Free BEC} along $x_3$ is
\begin{equation} \label{eq:BECfit}
    \Tilde{n}_0 = \frac{5}{2\pi}\frac{N_0}{R_1R_2}\cdot \Theta\left(1 - \sum_{i=1}^2\frac{x_i^2}{R_i^2}\right)^{\frac{3}{2}},
\end{equation}
where the explicit time-dependence of $R_i$ has been omitted.

It must be assumed that $\Tilde{z}=1$ for $T<T_c$, where $T_c$ is the critical temperature of a trapped ideal gas~\cite{Tc}, to obtain an analytical approximation for the thermal distribution, since a positive value for $\mu$, as predicted by the Thomas-Fermi expression below $T_c$, would cause the polylogarithm function to diverge. Thus, $\mu$ can only be included by using semi-ideal model, as discussed in the next section, which does not have a known analytic expression for the time-of-flight expansion. By integrating Eq.~\eqref{eq. time dependent thermal cloud} along $x_3$ and normalizing the expression to the number of thermal atoms $N_{th}$, the thermal cloud column density is approximated as
\begin{equation} \label{eq:TCfit}
\Tilde{n}_{th} = \frac{N_{th}}{2\pi w_1w_2}\frac{1}{g_3(1)}\cdot g_2\left(\e^{-\sum_{i=1}^2x_i^2/2w_i^2}\right).
\end{equation}
These equations assume no interaction between the thermal and condensed parts of the cloud and can be used to fit experimental images of bimodal clouds after time-of-flight. 

In our experiments, such fits are performed as follows. In a region centered on the cloud outside an ellipse with the radii $1.2R_i$ and within another ellipse with radii $3w_i$, the thermal component of the cloud is fitted and then subtracted from the total density distribution, followed by a fit of the remaining atoms with Eq.~\eqref{eq:BECfit} inside the ellipse with radii $1.2R_i$. The fit is performed twice, first to estimate the size of the regions, and then to extract $N_{th}$, $T$, and $N_0$. The temperature is extracted from the fitted widths in each direction, providing two values $T_1$ and $T_2$ which are averaged to obtain $T=\half(T_1+T_2)$.

\section{Simulation of partially condensed cloud expansion}
\label{sec:ExpandingMultimodalCloud}

Due to their high density, the BEC atoms affect the distribution of thermal atoms. This can be modeled by a mean-field approximation of the $s$-wave scattering between atoms, which leads to extra terms in the potential energy~\cite{Naraschewski1998, Giorgini, Griffin}
\begin{align}
    V_0 &= V + 2U_0n_{th},\\
    V_{th} &= V+2U_0(n_{th}+n_0),
\end{align}
where $V_0$ is the potential experienced by the BEC atoms and $V_{th}$ is the potential experienced by the thermal atoms. The forces from the BEC atoms on the BEC itself are already incorporated in the Thomas-Fermi model. Here, $U_0 = 4\pi\hbar^2 a/m$ is the interaction energy. Since the density of thermal atoms is low, the term $2U_0n_{th}$ can be neglected from both effective potentials. The result is an unaffected BEC cloud, and a thermal cloud that is repelled from the trap center by the BEC.

\begin{figure}[t]
    \centering\includegraphics[width=0.4\linewidth]{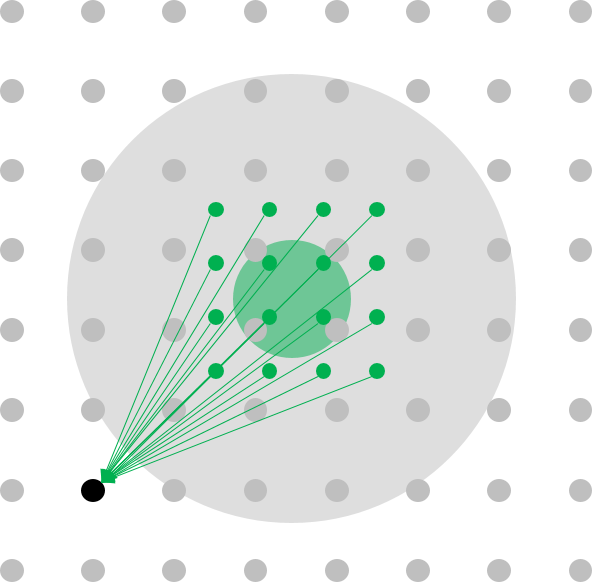}
    \caption{Illustration of the calculation of the thermal cloud profile after time-of-flight. The initial cloud, as well as the initial grid, is illustrated in green (small). The final cloud, as well as the expansion grid, is shown in grey (large). A black point in the expansion grid is highlighted. To calculate the density of atoms at this point, the probability of starting at a specific initial grid point, with the correct momentum to end up at the black point, is calculated for each initial grid point and then added. This is done for all expansion grid points to obtain the thermal atom distribution. The grids and clouds shown here are only an illustration of the method. The clouds are not assumed to be spherical, and the calculations are performed in three dimensions.}
    \label{fig:Caclulation Method}
\end{figure}

\begin{figure}[t]
    \centering\includegraphics[width=0.6\linewidth]{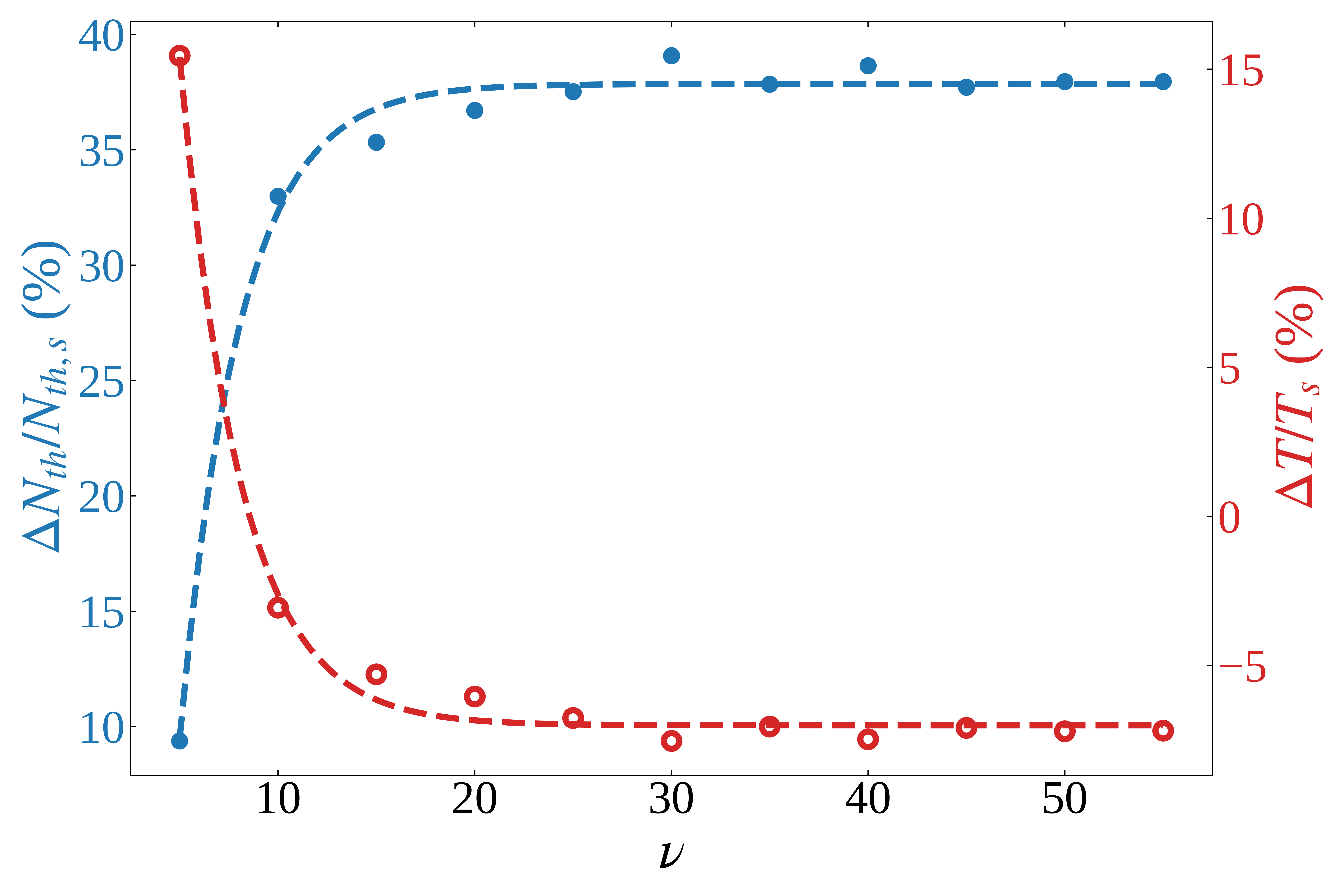}
    \caption{Relative errors in thermal atom number ({\color{bluemarkercolor}$\bullet$}) and temperature ({\color{redmarkercolor}\cir{}}) as a function of the grid size $\nu$ with $\xi=60$ held constant and for the cloud parameters specified in text. Exponential fits are provided as a guide to the eye.}
    \label{fig:spatialfinesse}
\end{figure}

\begin{figure}[t]
    \centering\includegraphics[width=0.6\linewidth]{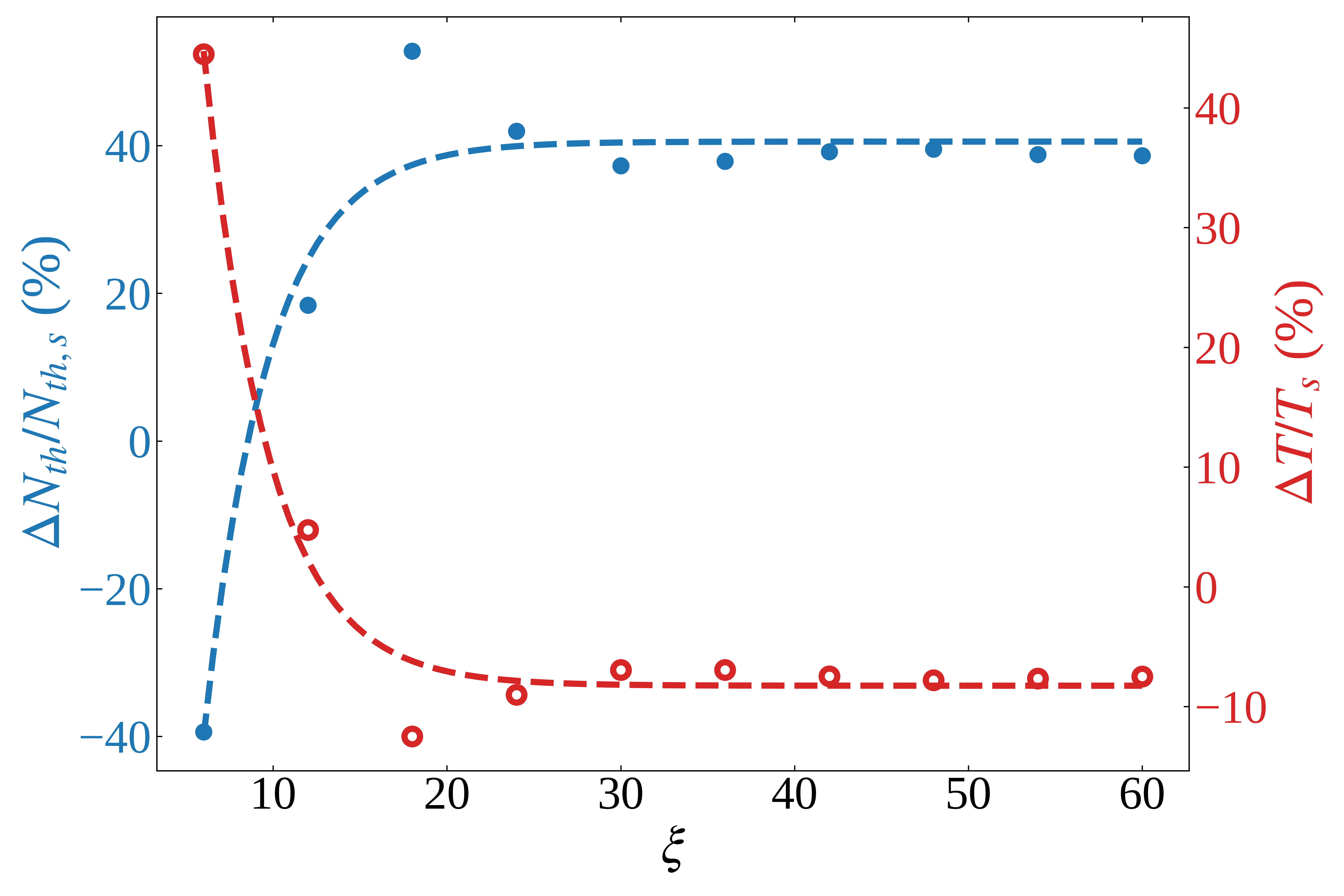}
    \caption{Relative errors in thermal atom number ({\color{bluemarkercolor}$\bullet$}) and temperature ({\color{redmarkercolor}\cir{}}) as a function of the grid size $\xi$ with $\nu=40$ held constant and for the cloud parameters specified in text. Exponential fits are provided as a guide to the eye.}
    \label{fig:momentumfinesse}
\end{figure}

\begin{figure}[t]
    \centering\includegraphics[width=0.6\linewidth]{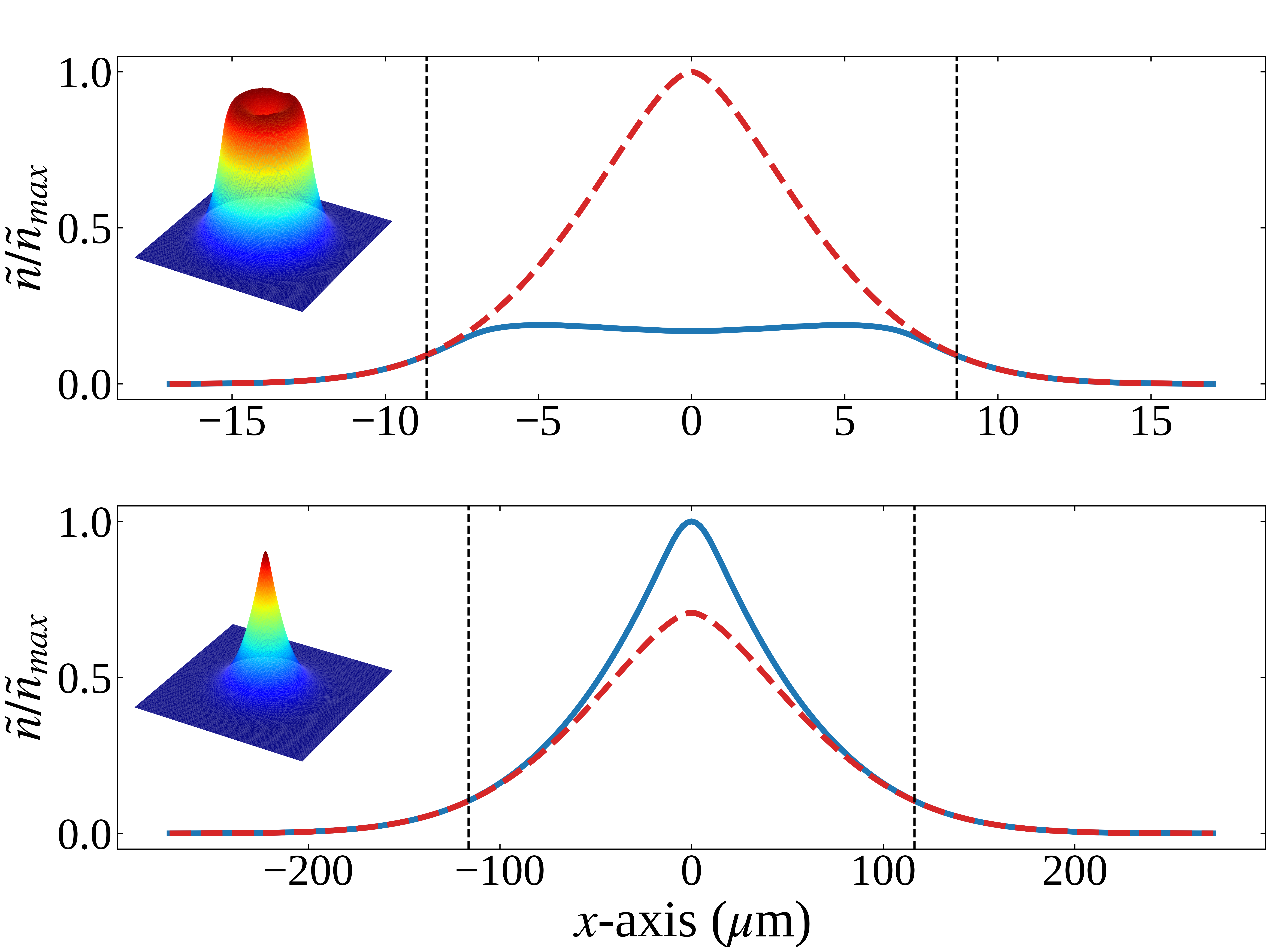}
    \caption{Thermal column densities in a spherical trap configuration with $N_s=\SI{200}{k}$, $T_s/T_{c,s}=0.2$, $\omega_i=2\pi\cdot\SI{100}{Hz}$, and $N_{0,s}/N_s=0.97$, for two different times of flight, normalized to the highest column density. The $x$-axis has been rescaled between the two plots. The BEC distribution and its fit are not plotted. The top has a time of flight of $t= \SI{1}{ms}$ and the bottom has $t= \SI{30}{ms}$. Slices through the center of the simulated column densities are shown as solid blue lines, and slices through the center of a Bose-enhanced fit to the simulated data (fitted outside the BEC region illustrated by vertical dashed lines) are shown as dashed red lines. Inserts in both figures show the simulated column densities.}
    \label{fig:3DPlusCut}
\end{figure}

In the following, the expansion of a released thermal cloud 
is found while ignoring the interactions after the trap is turned off, resulting in ballistic expansion. In reality, the interactions between the two components will quickly decrease as they expand rather than suddenly disappear. 

The time-dependent thermal cloud profile is again given by integrating over momentum space
\begin{align}
    \nonumber n_{th}(\textbf{r},t) &= \int\frac{\diff^3 \textbf{p}}{(2\pi\hbar)^3}\frac{1}{\e^{(p^2/2m + V_{\text{eff}}(\textbf{r},\textbf{p},t))/k_BT}-1},\\
    \nonumber V_{\text{eff}}(\textbf{r},\textbf{p},t)&=V(\textbf{r}-\textbf{p}/m\cdot t) + 2U_0n_0(\textbf{r}-\textbf{p}/m\cdot t) - \mu\\
    &=|\mu - V(\textbf{r}-\textbf{p}/m\cdot t)|\label{eq: final intrgral},
\end{align}
but, unlike Eq.~\eqref{eq. time dependent thermal cloud}, this integral cannot be solved analytically.

Cloud parameters used in the simulation of the expansion will be indicated with a subscript $s$. We numerically calculate the distribution after time-of-flight as follows. First, the physical parameters $N_s$, $T_s$, and a time-of-flight $t$ are chosen, where $N_s=N_{0,s}+N_{th,s}$ is the total number of atoms, for a trap with frequencies $\omega_i$. Then, a $\nu\cross\nu\cross\nu$ expansion grid for the calculation of the cloud profile after time-of-flight is defined, where $\nu$ is a positive integer. The range of the grid is defined to be four times the width of the thermal cloud after time-of-flight, calculated with the non-interacting model. Moreover, a $\xi\cross\xi\cross\xi$ grid around the in-trap cloud at $t=0$ is defined, called the initial grid, where $\xi$ is also a positive integer, with the range given by four times the in-trap width of the cloud.

The momentum $\mathbf{p}$ in Eq.~\eqref{eq: final intrgral} for an atom to move from a point in the initial grid, $\mathbf{r}_0$, to a point in the expansion grid, $\mathbf{r}$, during time-of-flight $t$ is given by $\mathbf{p}=m(\mathbf{r} - \mathbf{r}_0)/t$. The integral is evaluated numerically with a sum over the initial grid for each point in the expansion grid. This method is illustrated in FIG.~\ref{fig:Caclulation Method}. Finally, the grid points along $x_3$ are summed to calculate the column density $\Tilde{n}$.

The necessary grid sizes $\nu$ and $\xi$ are found by performing the calculations with varying sizes for the cloud that sets the highest numerical demands. This cloud corresponds to the parameters $N_s=10^6$, $T_s=0.3~T_{c,s}$, $t=\SI{3}{ms}$, $\omega_x=2\pi\cdot\SI{10}{Hz}$, and $\omega_{y,z}=2\pi\cdot\SI{100}{Hz}$. The simulated thermal distribution andthe theoretical BEC distribution after time-of-flight are added to obtain the total distribution. This distribution is fitted with the algorithm described in Section~\ref{subsec:ColumnDensities}, and the extracted thermal atom number $N_{th}$ and temperature $T$ are compared to the input values of the simulation, resulting in the errors $\Delta N_{th}=N_{th}-N_{th,s}$ and $\Delta T=T-T_{s}$, where the subscript $s$ indicates the values of the simulation. The relative errors as a function of the grid size $\nu$ are shown in FIG.~\ref{fig:spatialfinesse} with $\xi=60$ held constant. Similarly, FIG.~\ref{fig:momentumfinesse} displays the relative errors as a function of the initial grid size with $\nu=40$ held constant. Importantly, significant errors in both atom number and temperature are observed when fitting the results of the simulation with functions for non-interacting clouds from Sec.~\ref{subsec:ColumnDensities}. This is discussed in detail in the following sections. However, both figures show clear convergence of the estimation errors for sufficiently large grid sizes, and in the rest of this work $\nu=40$ and $\xi=60$ are used. Both of these choices can be relaxed if optimizing for speed. The complexity of the calculations scales as $\mathcal{O}(\nu^3\cdot \xi^3)$. Therefore, calculations are performed in parallel on a GeForce GTX 1050 Mobile graphics card.

To reduce the complexity of the simulation, cylindrical symmetry is assumed, since at least two out of the three trap frequencies are often equal. This means the spatial cloud profile after time-of-flight can be calculated in two dimensions, and the rotational symmetry can be exploited to obtain the 3-dimensional distribution. This assumption is used here, and it is often applicable to experimental realizations of BEC. There are three categories of cloud shapes within this assumption: a cigar-shaped cloud~\cite{Gorlitz2001, Schweikhard2004, S.Burger_2002, Hadzibabic2004, Andrews1997, davidClementCigar} if the two equal frequencies are larger than the remaining frequency, a pancake-shaped cloud if the two equal frequencies are lower than the remaining frequency~\cite{Rychtarik2004, Hammes2002, Smith_2005}, and a spherical cloud if all frequencies are the same. Spherical clouds can be simulated even more efficiently by exploiting the rotational symmetry around both axes. In all of these cases, it is only necessary to calculate the distribution in the positive quadrant ($x_i\ge 0$, $i=1,2,3$). 

The results of such a simulation are shown in FIG.~\ref{fig:3DPlusCut}. The figure shows cuts through the center of the simulated thermal cloud column densities, while the BEC density is not shown. The repulsive effect of BEC on the thermal cloud is most clearly seen at short time-of-flight (top), where a dip in the thermal column density in the center occurs. In addition, the fitted Bose-enhanced column densities according to the algorithm described in Sec.~\ref{subsec:ColumnDensities} are shown. While these fits agree well in the wings, they disagree significantly in the center. For a time of flight of $\SI{1}{ms}$, the fit overestimates the temperature by a relative error of 7.2\% due to the broad wings caused by the repulsive BEC. Similarly,  the atom number is overestimated by 64.5\%, since the thermal fit does not include the dip in density in the center. After $\SI{30}{ms}$ time-of-flight, some of the thermal atoms have filled the hole in the middle, and the cloud now has a slightly higher peak column density than the Bose-enhanced distribution. In this case, the relative error of the fit is -1.3\% for temperature and -16.9\% for atom number. Note that the sign of the error has reversed in this case because the simulated distribution is now more peaked than the Bose-enhanced distribution. 

\section{Atom Number and Temperature Errors}
\label{sec:Errors}

The analysis above motivates the error estimates for a broad range of parameters in this section. As described above, the distribution after time-of-flight is calculated for specific values of $N_s=N_{0,s}+N_{\text{th,s}}$ and $T_{s}$ and fitted according to Section~\ref{subsec:ColumnDensities} to extract $N_{th}$ and $T$. The errors are again given by $\Delta N_{th}=N_{th}-N_{th,s}$ and $\Delta T=T-T_{s}$.

\begin{figure}[t]
    \centering\includegraphics[width=1\linewidth]{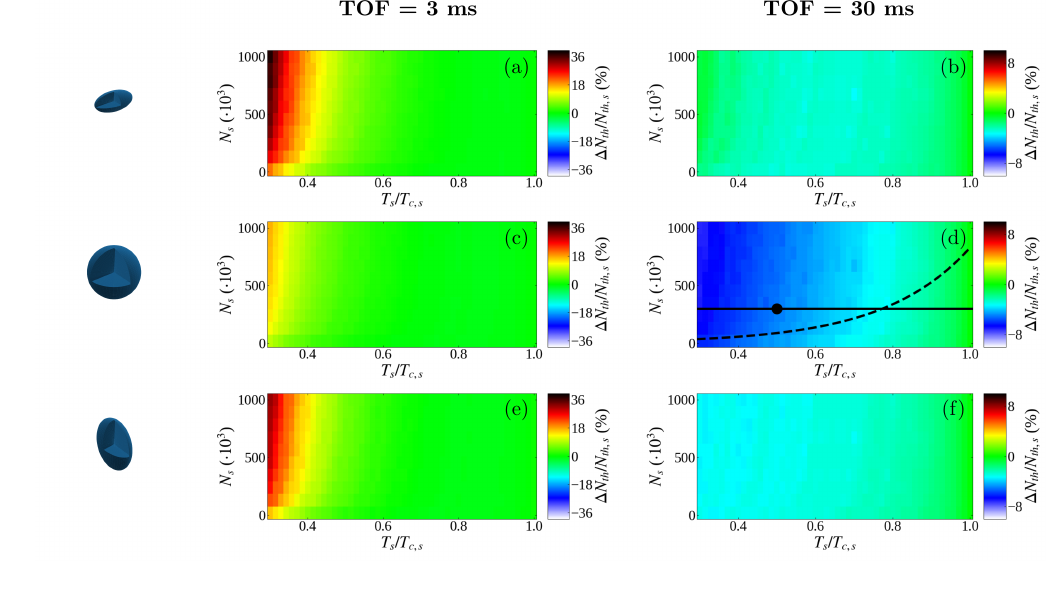}
    \caption{Errors in thermal atom number estimation as a function of reduced temperature $T_s/T_{c,s}$ and total atom number for different trap configurations and times of flight. (a) Cigar shape: $\omega_x = 2\pi\cdot\SI{10}{Hz}$, $\omega_{y,z} = 2\pi\cdot\SI{100}{Hz}$ and $t=\SI{3}{ms}$. (b) Cigar shape: $t=\SI{30}{ms}$. (c) Spherical shape: $\omega_{x,y,z} = 2\pi\cdot\SI{100}{Hz}$ and $t=\SI{3}{ms}$. (d) Spherical shape: $t=\SI{30}{ms}$. (e) Pancake shape: $\omega_x = 2\pi\cdot\SI{1000}{Hz}$, $\omega_{y,z} = 2\pi\cdot\SI{100}{Hz}$ and $t=\SI{3}{ms}$. (f) Pancake shape: $t=\SI{30}{ms}$. A solid line, a dashed line, and a solid circle indicate regions further explored in FIGs.~\ref{fig:EvapConstN},~\ref{fig:TOFScan},~and~\ref{fig:aspRatio}. }
    \label{fig:2DN}
\end{figure}

Figure~\ref{fig:2DN} shows the error in the thermal atom number determination for various cloud and trap parameters. The relative error $\Delta N_{th}/N_{th,s}$ is given for for three different trap geometries (cigar shaped trap with $\omega_x = \SI{10}{Hz}$, $\omega_{y,z} = \SI{100}{Hz}$; spherical trap with $\omega_{x,y,z} = \SI{100}{Hz}$; pancake shaped trap $\omega_x = \SI{1000}{Hz}$, $\omega_{y,z} = \SI{100}{Hz}$) and two different times-of-flight ($t=\SI{3}{ms}$; $t=\SI{30}{ms}$) as a function of the total atom number and the reduced temperature $T_s/T_{c,s}$. Positive values of the error indicate an overestimation by the Bose-enhanced fit function, Eq.~\eqref{eq:TCfit}. 

The two columns in the Fig.~\ref{fig:2DN} illustrate the difference between short (left) and long (right) time-of-flight. A longer time-of-flight generally reduces the magnitude of the errors; however, they do not approach zero. The sign of the errors reverses from short to long time-of-flight since the thermal cloud transitions from more flattened to more peaked compared to a Bose-enhanced distribution, as seen in FIG.~\ref{fig:3DPlusCut}.

The errors in thermal atom numbers are weakly dependent on atom number since the width of the thermal distribution is independent of the number of atoms, while the BEC only grows slowly in size with more atoms, since the Thomas-Fermi radius scales proportionally to $N_0^{1/5}$. Conversely, the temperature is more directly related to the size of the BEC and therefore has a much larger effect. Additionally, falling temperatures decrease the thermal cloud width while increasing the overlap between the two components. At $T_s=T_{c,s}$, the error is always zero as there are no condensed atoms, and the Bose-enhanced fit captures the simulation perfectly. 

The area surrounding the BEC is excluded from the fit of the thermal distribution. Consequently, a larger BEC region results in a less accurate estimate of the number of atoms in the thermal distribution. At long time-of-flight, the size of the BEC scales inversely to its size in-trap. 
For asymmetric trapping potentials, the direction of the smallest trapping frequency typically shows little expansion, thus allowing for a good atom number determination. However, for a cloud in a spherical trap, imaged after a certain time-of-flight, there is typically a temperature (equivalent to a chemical potential) at which the BEC approaches the thermal cloud size. This can explain the patterns shown in FIG.~\ref{fig:2DN} at long time-of-flight, where the errors in thermal atom numbers in spherical clouds are larger than in both asymmetrical clouds.

\begin{figure}[t]
    \centering\includegraphics[width=1\linewidth]{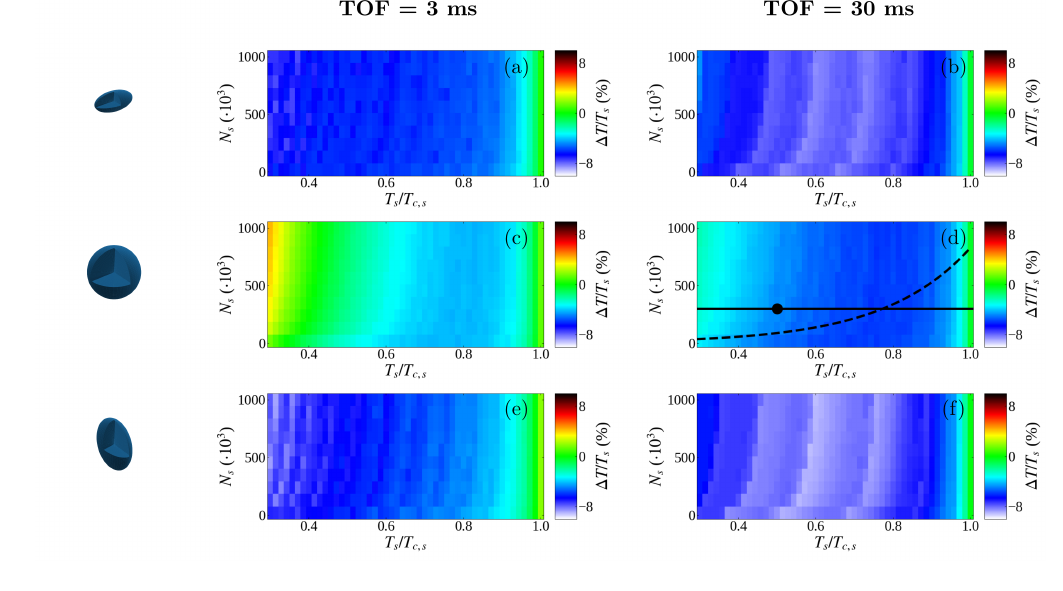}
    \caption{Errors in temperature estimation as a function of reduced temperature $T_s/T_{c,s}$ and total atom number for different trap configurations and times of flight. The parameters are the same as in FIG.~\ref{fig:2DN}, specifically: (a) Cigar shape: $\omega_x = 2\pi\cdot\SI{10}{Hz}$, $\omega_{y,z} = 2\pi\cdot\SI{100}{Hz}$ and $t=\SI{3}{ms}$. (b) Cigar shape: $t=\SI{30}{ms}$. (c) Spherical shape: $\omega_{x,y,z} = 2\pi\cdot\SI{100}{Hz}$ and $t=\SI{3}{ms}$. (d) Spherical shape: $t=\SI{30}{ms}$. (e) Pancake shape: $\omega_x = 2\pi\cdot\SI{1000}{Hz}$, $\omega_{y,z} = 2\pi\cdot\SI{100}{Hz}$ and $t=\SI{3}{ms}$. (f) Pancake shape: $t=\SI{30}{ms}$. A solid line, a dashed line, and a solid circle indicate regions further explored in FIGs.~\ref{fig:EvapConstN},~\ref{fig:TOFScan},~and~\ref{fig:aspRatio}. The clearly visible, almost vertical, stripes are an artifact of numerical uncertainty, and oscillate on a less than 1\% level in the estimated error in parameters. }
    \label{fig:2DTM}
\end{figure}

Figure~\ref{fig:2DTM} shows the errors in the temperature estimation for the same parameters as FIG.~\ref{fig:2DN}. For all parameters simulated here, the relative temperature errors are within $\pm$10\%. The temperature errors, like the errors on atom number, are weakly dependent on atom number, while the dependence on temperature is slightly more significant. A particularly interesting case is the spherical short time-of-flight example, which shows a reversal in the sign of the error at low temperatures. The reason is the same as the reversal in sign shown in FIG.~\ref{fig:3DPlusCut}. A time-of-flight of $t=\SI{3}{ms}$ is enough to reverse the sign of the error for moderately low temperatures, but for the coldest clouds, more time-of-flight is needed before the reversal occurs. Interestingly, this means there are regions with no errors. The same effect can also be seen in the spherical long time-of-flight plot, where $t=\SI{30}{ms}$ is not enough to fully reverse the error for the very coldest clouds. 

Overall, the pattern in temperature errors for different trap geometries is roughly opposite to that seen for thermal atom number errors. This can be understood based on the following argument. A smaller size of the BEC after expansion allows the thermal cloud fit function to access more of the central region where the cloud is deformed the most. While this provides a better estimate of the area under the fitted curve, and thus lower thermal atom number errors, it requires the fundamentally incorrect Bose-enhanced fit function to adapt to the cloud shape in the central region. Spherical clouds are thus the best for estimating cloud temperatures when neglecting the repulsive forces of the BEC, while they are the worst for estimating atom numbers.

\begin{figure}[t]
    \centering\includegraphics[width=0.6\linewidth]{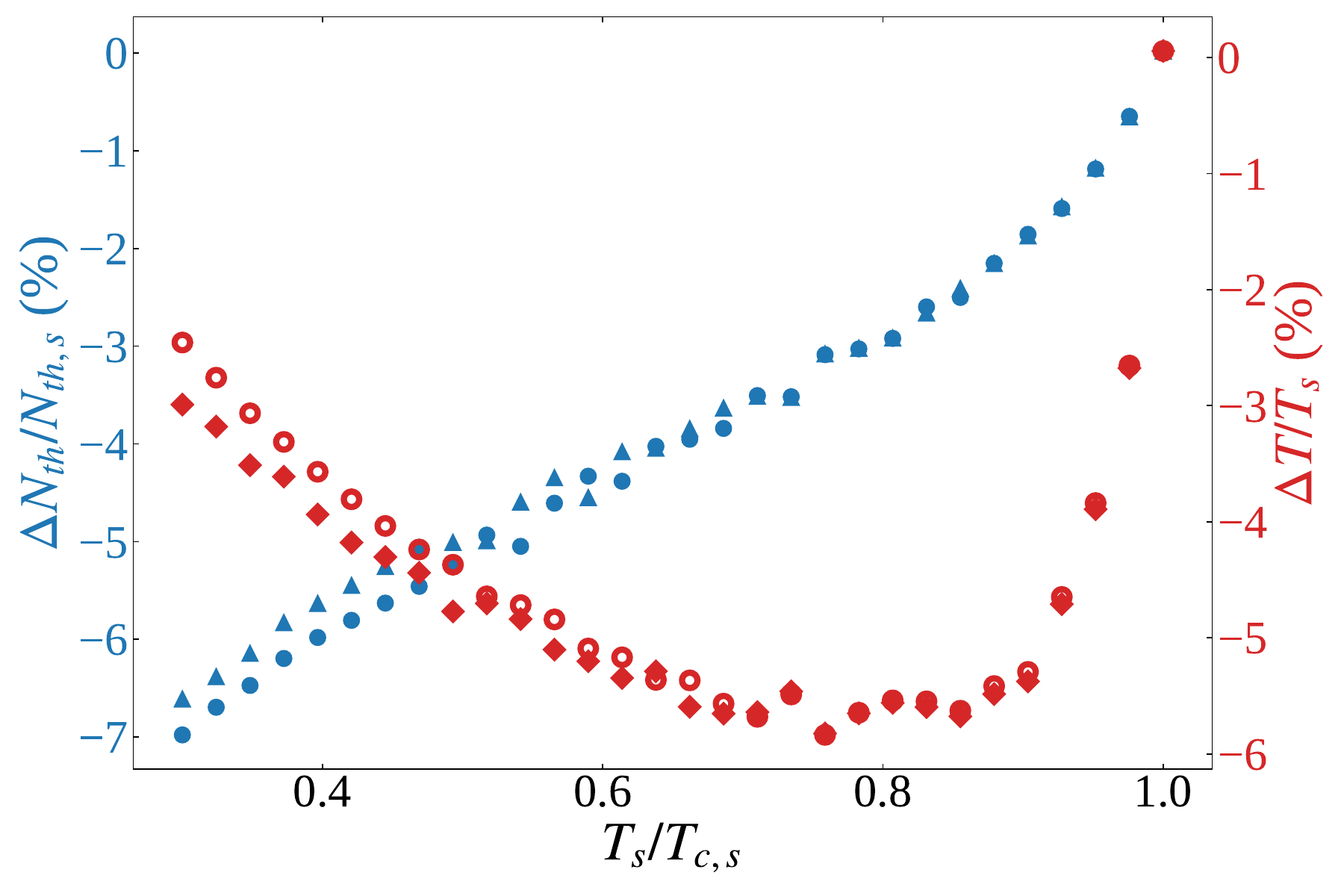}
    \caption{Errors in relative thermal atom numbers and temperature for different temperatures. The simulations show results for a spherical trap at $N_s=3\cdot 10^5$ and $t=\SI{30}{ms}$. The {\color{bluemarkercolor}$\bullet$}- and {\tiny\color{bluemarkercolor}\ding{115}}-markers show the thermal atom number error along a constant atom number and an experimental evaporative trajectory respectively, and the {\color{redmarkercolor}\cir{}}- and {\scriptsize\color{redmarkercolor}\ding{117}}-markers show the temperature errors along a constant atom number and an experimental evaporative trajectory respectively. Both trajectories are illustrated in FIG.~\ref{fig:2DTM}(d)).}
    \label{fig:EvapConstN}
\end{figure}

The errors in atom number and temperature were explored further for a selected range of parameters. The dashed lines in FIGs.~\ref{fig:2DN}(d)~and~\ref{fig:2DTM}(d) indicate a typical evaporative cooling trajectory that reduces both the atom number and temperature. For comparison, a hypothetical trajectory at a constant atom number of $3\cdot10^5$ (solid line) is also shown. Figure~\ref{fig:EvapConstN} shows the relative errors in atom number and temperature as a function of temperature in both cases. For both trajectories, the underestimation of the number of thermal atoms increases towards low temperatures, reaching $\approx7\%$. Similarly, the temperature is underestimated as the temperature is decreased; however, it shows a maximal underestimation of $\approx6\%$ for $T_s/T_{c,s}=0.8$ and then returns to a moderate underestimation of about $\approx3\%$ for the lowest investigated temperatures. Note, that the two trajectories do not lead to significant differences.

\begin{figure}[t]
    \centering\includegraphics[width=0.6\linewidth]{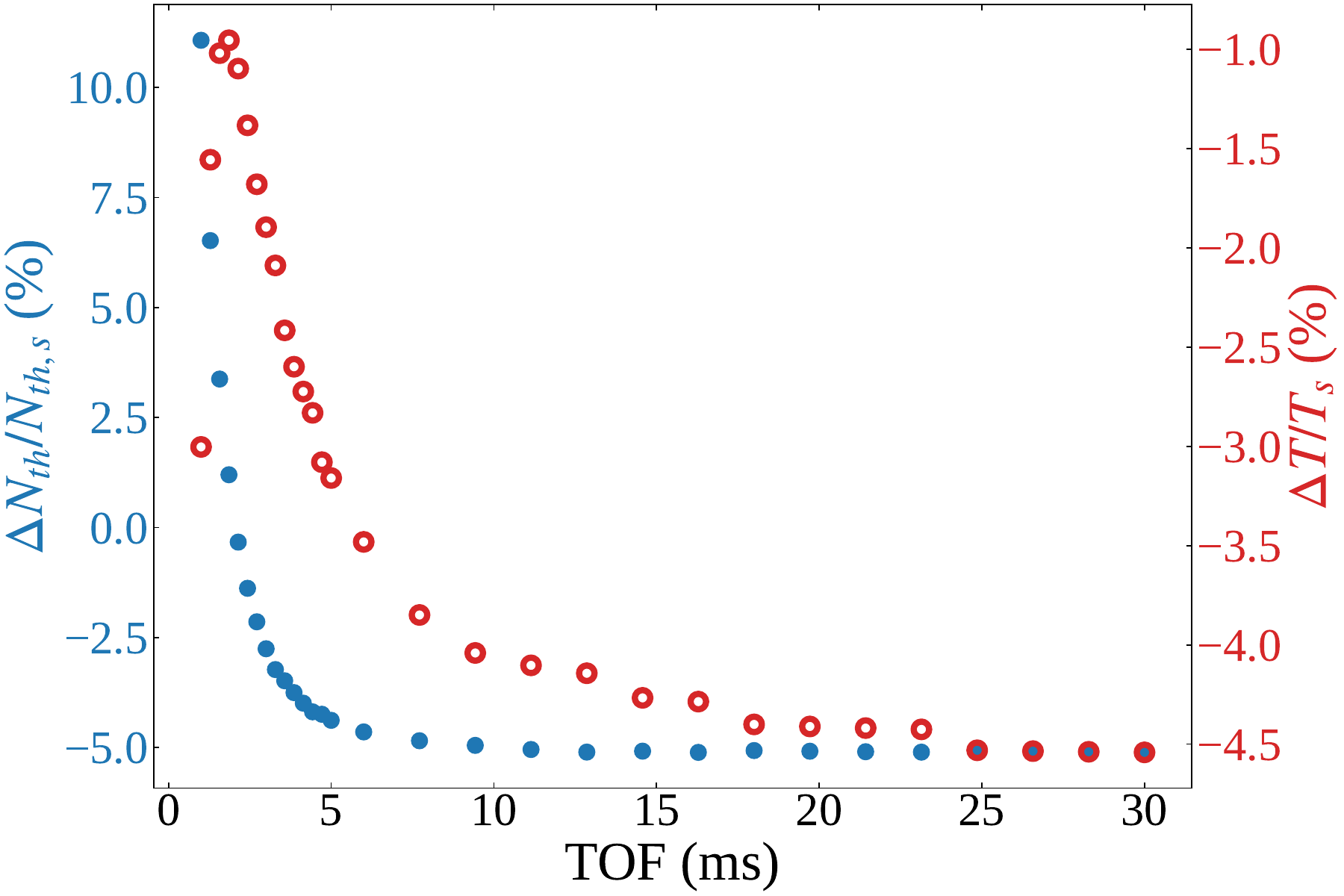}
    \caption{Relative errors in thermal atom number ({\color{bluemarkercolor}$\bullet$}) and temperature ({\color{redmarkercolor}\cir{}}) for different times of flight from a spherical trap. The simulations are performed at $N_s=3\cdot 10^5$ and $T_s/T_{c,s}=0.5$.}
    \label{fig:TOFScan}
\end{figure}

The dependence of the relative errors on the time of flight is explored in further detail in FIG.~\ref{fig:TOFScan} for the parameters indicated by a solid circle in FIGs.~\ref{fig:2DN}(d)~and~\ref{fig:2DTM}(d). The relative atom number error decreases quickly with expansion time, and converges to $\approx-5\%$. On the contrary, the relative temperature error first increases and then converges to $\approx-4.5\%$ on a slightly slower time scale. This is in agreement with the behavior seen in FIG.~\ref{fig:2DTM}(d), where the sign of the temperature error reverses for low temperatures, while in FIG.~\ref{fig:2DN}(d), the atom number error sign reversal has already occurred at $t=\SI{30}{ms}$. Importantly, the errors do not converge to zero for a long expansion times, since the BEC alters the energy levels populated by the thermal atoms and thus also the distribution of momenta. This affects the expansion of the cloud for all times of flight, resulting in a slower expansion and thus an underestimation of the temperature.

\begin{figure}[t]
    \centering\includegraphics[width=0.6\linewidth]{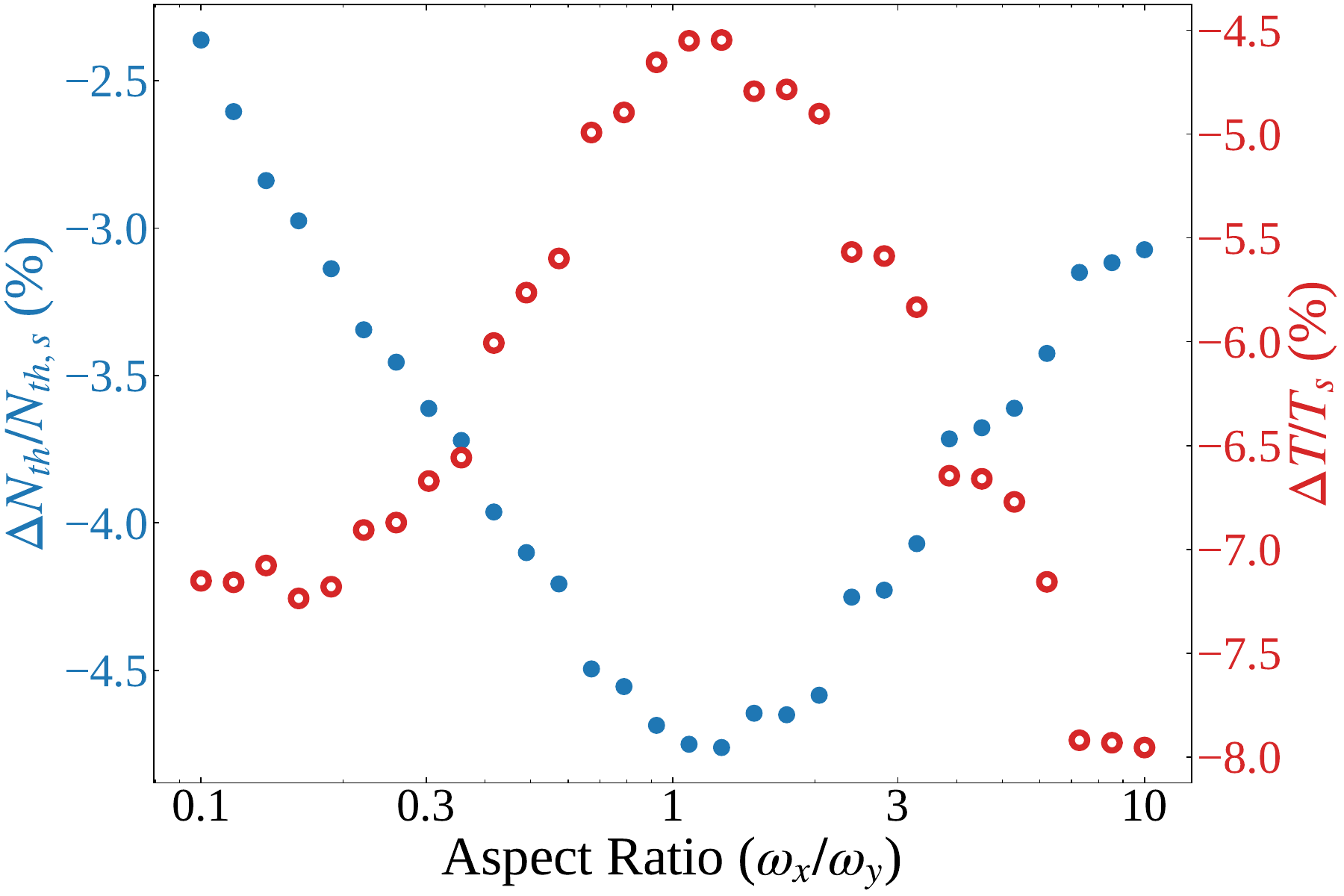}
    \caption{Errors in relative thermal atom number ({\color{bluemarkercolor}$\bullet$}) and temperature ({\color{redmarkercolor}\cir{}}) for different aspect ratios of the trap, but with a constant geometrical mean of the aspect ratios. The simulations are performed at $N_s=3\cdot 10^5$, $T_s/T_{c,s}=0.5$ and $t=\SI{30}{ms}$.}
    \label{fig:aspRatio}
\end{figure}

Finally, the effect of the trap geometry is investigated by varying the aspect ratio, at otherwise identical parameters as shown in FIG.~\ref{fig:aspRatio}. The errors in the thermal atom number are small at the two extremes and largest in the middle, where the cloud is spherical, while the opposite pattern emerges for the errors in temperature. This once again shows that the spherical cloud is best suited for temperature estimation but not for thermal atom number determination.

\section{Bimodal fits of partially condensed clouds}
\label{sec:Fits}

The simulation of the expansion can also be used within a fitting routine to obtain the thermal atom number and temperature more accurately from experimental data. The fit method follows the same steps as explained in Section~\ref{subsec:ColumnDensities}, with the difference that the thermal distribution is provided by the simulation. In the following, the performance of such a fit is compared to the fit with the Bose-enhanced function according to Eq.~\eqref{eq:TCfit}. Since the BEC and thermal components are fitted independently, the use of the semi-ideal model in the simulation of the cloud expansion does not guarantee that the BEC fraction follows the theoretically expected behavior as a function of temperature~\cite{Naraschewski1998}. 

The experimental apparatus to produce partially condensed BECs was previously described in detail~\cite{Christensen2020}. Briefly, \Rb{} atoms are initially cooled by radio-frequency (RF) evaporation in a Ioffe-Pritchard type magnetic trap. To stabilize the production of ultracold clouds, the cooling process is interrupted at a temperature of $\sim 14$~$\mu$K, and the atomic clouds are probed using minimally destructive Faraday imaging~\cite{Gajdacz2013,Kristensen2017}. Based on the outcome, the atom number is corrected by removing excess atoms. Subsequently, the magnetic trap is decompressed to obtain radial and axial trapping frequencies of  $\omega_\rho = 2\pi \times \SI{93.4}{\hertz} $ and $\omega_z = 2\pi \times \SI{17.7}{ \hertz}$. Finally, BECs are produced by applying an RF sweep that leads to the desired average BEC occupation. The partially condensed clouds are probed using absorption imaging after a \SI{27.5}{\milli\second} time of flight. To accurately detect the density distribution of dense clouds, the imaging system was calibrated following Refs.~\cite{img1,img4} and an intensity of $I=2.3~I_\mathrm{sat}$ was used. 

\begin{figure}
    \centering\includegraphics[width=0.6\linewidth]{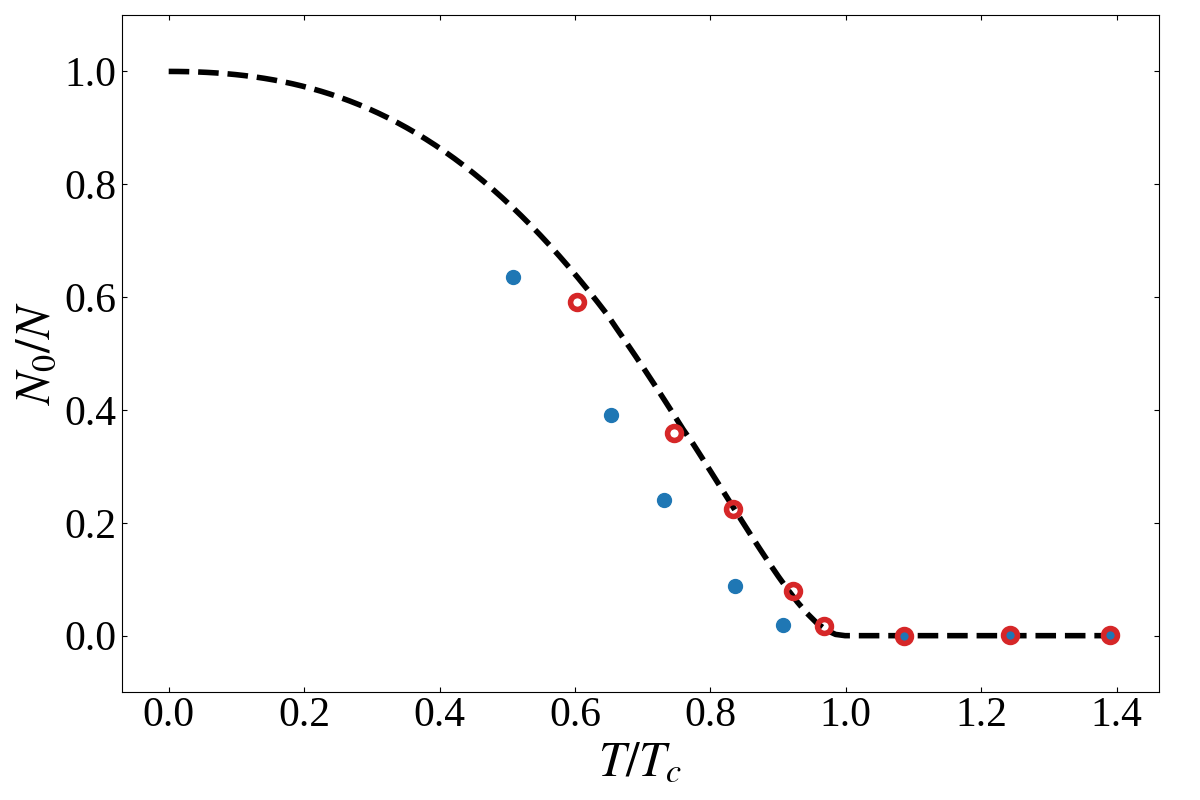}
    \caption{BEC fraction as a function of reduced temperature $T/T_c$. The results of an analytic fit with the Bose-enhanced distribution in Eq.~\eqref{eq:TCfit} ({\color{bluemarkercolor}$\bullet$}) and the results of the fit with the simulated distribution ({\color{redmarkercolor}\cir{}}) are compared to the semi-ideal model~\cite{Naraschewski1998}. Each point is an average of 6 to 8 fits. The uncertainties of the mean values are too small to see on this plot.}
    \label{fig:N0_T}
\end{figure}

A fit with the simulated distribution after time-of-flight takes about ten minutes, and thus, the fits were performed on a small dataset with six to eight images at five different final RF frequencies. The same images were also fitted with the Bose-enhanced function for comparison. Figure~\ref{fig:N0_T} shows the results of both fitting techniques, by displaying the extracted BEC fraction as a function of reduced temperature against the semi-ideal model~\cite{Naraschewski1998}. The fits with simulated clouds are in excellent agreement with the results of the semi-ideal model, showing the improvement to be gained from this technique. In comparison, the Bose-enhanced fit function, which completely ignores the repulsive forces of the BEC, underestimates the temperature significantly while overestimating the BEC fraction slightly. Note, however, that even the simulation ignores interactions during the time-of-flight, which might explain the small remaining discrepancies. 

\section{Conclusion}
\label{sec:Conclusion}

The repulsive interactions between the condensed and thermal atoms in a partially condensed BEC reshape the thermal distribution and change the expansion profile of the cloud during time-of-flight. This results in errors in the estimation of temperature and atom number if not taken into account. Since the expansion of partially condensed clouds can not be calculated analytically, numerical simulations of this expansion are performed. In the simulations presented here, the interactions are taken into account in the trap while assuming ballistic expansion during time-of-flight. To mimic the experimental situation, these simulations are then fitted with a Bose-enhanced function to obtain atom numbers and temperatures. The difference between the input values of the simulation and the extracted parameters provides an estimate of the errors typically made in experimental realizations.

Our simulations show that these errors are significant even for long times of flight and in general depend on cloud temperature, trap geometry, and more weakly on the total number of atoms in the cloud. We provide a phenomenological explanation for the patterns observed in the errors. The size of the BEC after expansion determines the area where the thermal cloud can be fitted. Estimation of thermal atom numbers using a Bose-enhanced fit function generally benefits from a larger fit region, while access to a fit region close to the cloud centre can lead to significant temperature errors. Consequently, there is a trade-off in the estimation of these two parameters, which depends on the time of flight and the trapping configuration. Our method can be used to provide systematic error estimates for experiments in which the Bose-enhanced fit function is used. 

In addition, our simulation can be used as a fitting model for experimental data. By applying such a fit to experimental images of ultracold clouds in the vicinity of the BEC transition, it was shown that the extracted parameters are in considerably better agreement with the semi-ideal model~\cite{Naraschewski1998} than a Bose-enhanced fit. The fit of each cloud with the simulation took about ten minutes on a single graphics card. It is thus much slower than the Bose-enhanced fit, but with appropriate hardware, it can be used to analyze experiments in real time. Importantly, the model can therefore be used to achieve higher precision cloud parameters in experiments with partially condensed clouds. 

The simulations did not take interactions during the time-of-flight into account. These effects will be investigated in future work, but it will require a fundamentally different approach, since the ballistic expansion model is no longer applicable. 

\section*{Acknowledgments}
\label{sec:Acknowledgments}
The authors thank Adam Simon Chatterley for valuable discussions. We acknowledge support from the Danish National Research Foundation through the Center of Excellence ``CCQ'' (DNRF152) and by the Novo Nordisk Foundation NERD grant (Grant number NNF22OC0075986). C.W. acknowledges support from EPSRC (Grant no. EP/Y004728/1).

\bibliography{bib}

@Article{CastinDum,
  author    = {Y. Castin and R. Dum},
  journal   = {Phys. Rev. Lett.},
  title     = {{Bose}-{Einstein} Condensates in Time Dependent Traps},
  year      = {1996},
  month     = dec,
  number    = {27},
  pages     = {5315},
  volume    = {77},
  doi       = {https://doi.org/10.1103/PhysRevLett.77.5315},
  publisher = {American Physical Society ({APS})},
}

@Article{Griffin,
  author    = {A. Griffin},
  journal   = {Phys. Rev. B},
  title     = {Conserving and gapless approximations for an inhomogeneous {Bose} gas at finite temperatures},
  year      = {1996},
  month     = apr,
  number    = {14},
  pages     = {9341},
  volume    = {53},
  doi       = {https://doi.org/10.1103/PhysRevB.53.9341},
  publisher = {American Physical Society ({APS})},
}

@article{davidClementCigar,
  title = {Non-Gaussian Correlations in the Steady State of Driven-Dissipative Clouds of Two-Level Atoms},
  author = {Ferioli, Giovanni and Pancaldi, Sara and Glicenstein, Antoine and Cl\'ement, David and Browaeys, Antoine and Ferrier-Barbut, Igor},
  journal = {Phys. Rev. Lett.},
  volume = {132},
  issue = {13},
  pages = {133601},
  numpages = {6},
  year = {2024},
  month = {Mar},
  publisher = {American Physical Society},
  doi = {10.1103/PhysRevLett.132.133601},
  url = {https://link.aps.org/doi/10.1103/PhysRevLett.132.133601}
}

@Article{Giorgini,
  author    = {S. Giorgini and L. P. Pitaevskii and S. Stringari},
  journal   = {Phys. Rev. Lett.},
  title     = {Scaling and Thermodynamics of a Trapped {Bose}-Condensed Gas},
  year      = {1997},
  month     = may,
  number    = {21},
  pages     = {3987},
  volume    = {78},
  doi       = {https://doi.org/10.1103/PhysRevLett.78.3987},
  publisher = {American Physical Society ({APS})},
}

@PhdThesis{Christensen2020,
	author = {Mikkel Berg Christensen},
	school = {Aarhus University},
	title  = {Microcanonical Fluctuations in Interacting {Bose}-{Einstein} Condensates},
	year   = {2020},
	month  = {jul},
}

@PhdThesis{MalthePhD,
	author = {M. Andersen},
	school = {Aarhus University},
	title  = {Spectrally Probing Ultra-Cold Bosonic Clouds},
	year   = {2025},
	month  = Sep,
	url    = {LINK-LINK-LINK-LINK-LINK-LINK-LINK-LINK},
}

@article{Kristensen2017,
	doi = {10.1088/1361-6455/50/3/034004},
	url = {https://dx.doi.org/10.1088/1361-6455/50/3/034004},
	year = {2017},
	month = {jan},
	publisher = {IOP Publishing},
	volume = {50},
	number = {3},
	pages = {034004},
	author = {Kristensen, M A and Gajdacz, M and Pedersen, P L and Klempt, C and Sherson, J F and Arlt, J J and Hilliard, A J},
	title = {Sub-atom shot noise {Faraday} imaging of ultracold atom clouds},
	journal = {Journal of Physics B: Atomic, Molecular and Optical Physics}
}

@article{Gajdacz2013,
	author = {Gajdacz, Miroslav and Pedersen, Poul L. and Mørch, Troels and Hilliard, Andrew J. and Arlt, Jan and Sherson, Jacob F.},
	title = {Non-destructive {Faraday} imaging of dynamically controlled ultracold atoms},
	journal = {Review of Scientific Instruments},
	volume = {84},
	number = {8},
	pages = {083105},
	year = {2013},
	month = {08},
	issn = {0034-6748},
	doi = {10.1063/1.4818913},
	url = {https://doi.org/10.1063/1.4818913},
	eprint = {https://pubs.aip.org/aip/rsi/article-pdf/doi/10.1063/1.4818913/14781030/083105\_1\_online.pdf},
}

@Article{Naraschewski1998,
  author    = {M. Naraschewski and D. M. Stamper-Kurn},
  journal   = {Phys. Rev. A},
  title     = {Analytical description of a trapped semi-ideal {Bose} gas at finite temperature},
  year      = {1998},
  month     = sep,
  number    = {3},
  pages     = {2423},
  volume    = {58},
  doi       = {https://doi.org/10.1103/PhysRevA.58.2423},
  publisher = {American Physical Society ({APS})},
}

@article{Vibel,
doi = {10.1088/1361-6455/ad7458},
url = {https://dx.doi.org/10.1088/1361-6455/ad7458},
year = {2024},
month = {sep},
publisher = {IOP Publishing},
volume = {57},
number = {19},
pages = {195301},
author = {T Vibel and M B Christensen and R M F Andersen and L N Stokholm and K Pawłowski and K Rzążewski and M A Kristensen and J J Arlt},
title = {Atom number fluctuations in {Bose} gases—statistical analysis of parameter estimation},
journal = {Journal of Physics B: Atomic, Molecular and Optical Physics}
}

@article{thermalTOF,
	title = {Measurement of {Temperature} of {Atomic} {Cloud} {Using} {Time}-of-{Flight} {Technique}},
	volume = {27},
	issn = {0974-9853},
	url = {https://doi.org/10.1007/s12647-012-0003-3},
	doi = {10.1007/s12647-012-0003-3},
	number = {1},
	journal = {MAPAN},
	author = {Arora, P. and Purnapatra, S. B. and Acharya, A. and Kumar, R. and Sen Gupta, A.},
	month = mar,
	year = {2012},
	pages = {31--39},
}

@article{img1,
  title={Strong saturation absorption imaging of dense clouds of ultracold atoms},
  author={Reinaudi, G and Lahaye, T and Wang, Z and Gu{\'e}ry-Odelin, D},
  journal={Optics letters},
  volume={32},
  number={21},
  pages={3143--3145},
  year={2007},
  publisher={Optica Publishing Group}
}

@article{img2,
  title={Calibrating high intensity absorption imaging of ultracold atoms},
  author={Hueck, Klaus and Luick, Niclas and Sobirey, Lennart and Siegl, Jonas and Lompe, Thomas and Moritz, Henning and Clark, Logan W and Chin, Cheng},
  journal={Optics express},
  volume={25},
  number={8},
  pages={8670--8679},
  year={2017},
  publisher={Optica Publishing Group}
}

@article{img3,
  title={Quantitative absorption imaging: The role of incoherent multiple scattering in the saturating regime},
  author={Veyron, Romain and Mancois, Vincent and Gerent, Jean-Baptiste and Baclet, Guillaume and Bouyer, Philippe and Bernon, Simon},
  journal={Physical Review Research},
  volume={4},
  number={3},
  pages={033033},
  year={2022},
  publisher={APS}
}

@article{img4,
  title={Spatial calibration of high-density absorption imaging},
  author={Vibel, Toke and Christensen, Mikkel Berg and Kristensen, Mick Althoff and Thuesen, Jeppe Juhl and Stokholm, Laurits Nikolaj and Weidner, Carrie Ann and Arlt, Jan Joachim},
  journal={Journal of Physics B: Atomic, Molecular and Optical Physics},
  year={2024}
}

@article{img5,
  title={Analysis and calibration of absorptive images of {Bose}--{Einstein} condensate at nonzero temperatures},
  author={Szczepkowski, J and Gartman, R and Witkowski, M and Tracewski, L and Zawada, M and Gawlik, W},
  journal={Review of Scientific Instruments},
  volume={80},
  number={5},
  year={2009},
  publisher={AIP Publishing}
}

@incollection{img6,
  title={Making, probing and understanding {Bose}-{Einstein} condensates},
  author={Ketterle, Wolfgang and Durfee, Dallin S and Stamper-Kurn, DM},
  pages={67--176},
  year={1999},
  publisher={IOS Press}
}

@article{img7,
  title={Optimized absorption imaging of mesoscopic atomic clouds},
  author={Muessel, Wolfgang and Strobel, Helmut and Joos, Maxime and Nicklas, Eike and Stroescu, Ion and Tomkovi{\v{c}}, Ji{\v{r}}{\'\i} and Hume, David B and Oberthaler, Markus K},
  journal={Applied Physics B},
  volume={113},
  pages={69--73},
  year={2013},
  publisher={Springer}
}

@article{img8,
  title={Feshbach enhanced s-wave scattering of fermions: direct observation with optimized absorption imaging},
  author={Genkina, Dina and Aycock, LM and Stuhl, BK and Lu, Hsin-I and Williams, RA and Spielman, IB},
  journal={New Journal of Physics},
  volume={18},
  number={1},
  pages={013001},
  year={2015},
  publisher={IOP Publishing}
}

@article{img9,
  title={Optimally focused cold atom systems obtained using density-density correlations},
  author={Putra, Andika and Campbell, Daniel L and Price, Ryan M and De, Subhadeep and Spielman, IB},
  journal={Review of Scientific Instruments},
  volume={85},
  number={1},
  year={2014},
  publisher={AIP Publishing}
}

@article{expandingBEC2,
  title = {Evolution of a {Bose}-condensed gas under variations of the confining potential},
  author = {Kagan, Yu. and Surkov, E. L. and Shlyapnikov, G. V.},
  journal = {Phys. Rev. A},
  volume = {54},
  issue = {3},
  pages = {R1753--R1756},
  numpages = {0},
  year = {1996},
  month = {Sep},
  publisher = {American Physical Society},
  doi = {10.1103/PhysRevA.54.R1753},
  url = {https://link.aps.org/doi/10.1103/PhysRevA.54.R1753}
}

@Book{Schroeder2014,
	author    = {Daniel V. Schroeder},
	publisher = {Pearson},
	title     = {An Introduction to Thermal Physics},
	year      = {2014},
	edition   = {First},
}

@Book{Pethick2008,
	author    = {Pethick, Christopher J. and Smith, Henrik},
	publisher = {Cambridge},
	title     = {{Bose}-{Einstein} condensation in dilute gases},
    edition   = {Second},
	year      = {2008},
}

@article{Gross1961,
	author       = {Eugene P. Gross},
	title        = {Structure of a quantized vortex in boson systems},
	journal      = {Il Nuovo Cimento},
	volume       = {20},
	pages        = {454--477},
	year         = {1961},
	month        = {may},
	doi          = {10.1007/BF02731494},
}

@article{Pitaevskii1961,
	author       = {Lev P. Pitaevskii},
	title        = {Vortex lines in an imperfect {Bose} gas},
	journal      = {Soviet Physics JETP},
	volume       = {13},
	number       = {2},
	pages        = {451--454},
	year         = {1961},
	month        = {aug},
}

@article{Rychtarik2004,
  title = {Two-Dimensional Bose-Einstein Condensate in an Optical Surface Trap},
  author = {Rychtarik, D. and Engeser, B. and N\"agerl, H.-C. and Grimm, R.},
  journal = {Phys. Rev. Lett.},
  volume = {92},
  issue = {17},
  pages = {173003},
  numpages = {4},
  year = {2004},
  month = {Apr},
  publisher = {American Physical Society},
  doi = {10.1103/PhysRevLett.92.173003},
  url = {https://link.aps.org/doi/10.1103/PhysRevLett.92.173003}
}

@article{Hammes2002,
  title = {Cold-atom gas at very high densities in an optical surface microtrap},
  author = {Hammes, M. and Rychtarik, D. and N\"agerl, H.-C. and Grimm, R.},
  journal = {Phys. Rev. A},
  volume = {66},
  issue = {5},
  pages = {051401},
  numpages = {4},
  year = {2002},
  month = {Nov},
  publisher = {American Physical Society},
  doi = {10.1103/PhysRevA.66.051401},
  url = {https://link.aps.org/doi/10.1103/PhysRevA.66.051401}
}

@article{Gorlitz2001,
  title = {Realization of Bose-Einstein Condensates in Lower Dimensions},
  author = {G\"orlitz, A. and Vogels, J. M. and Leanhardt, A. E. and Raman, C. and Gustavson, T. L. and Abo-Shaeer, J. R. and Chikkatur, A. P. and Gupta, S. and Inouye, S. and Rosenband, T. and Ketterle, W.},
  journal = {Phys. Rev. Lett.},
  volume = {87},
  issue = {13},
  pages = {130402},
  numpages = {4},
  year = {2001},
  month = {Sep},
  publisher = {American Physical Society},
  doi = {10.1103/PhysRevLett.87.130402},
  url = {https://link.aps.org/doi/10.1103/PhysRevLett.87.130402}
}

@article{Schweikhard2004,
  title = {Rapidly Rotating Bose-Einstein Condensates in and near the Lowest Landau Level},
  author = {Schweikhard, V. and Coddington, I. and Engels, P. and Mogendorff, V. P. and Cornell, E. A.},
  journal = {Phys. Rev. Lett.},
  volume = {92},
  issue = {4},
  pages = {040404},
  numpages = {4},
  year = {2004},
  month = {Jan},
  publisher = {American Physical Society},
  doi = {10.1103/PhysRevLett.92.040404},
  url = {https://link.aps.org/doi/10.1103/PhysRevLett.92.040404}
}

@article{Smith_2005,
doi = {10.1088/0953-4075/38/3/007},
url = {https://doi.org/10.1088/0953-4075/38/3/007},
year = {2005},
month = {jan},
publisher = {},
volume = {38},
number = {3},
pages = {223},
author = {Smith, N L and Heathcote, W H and Hechenblaikner, G and Nugent, E and Foot, C J},
title = {Quasi-2D confinement of a BEC in a combined optical and magnetic potential},
journal = {Journal of Physics B: Atomic, Molecular and Optical Physics}
}

@article{S.Burger_2002,
doi = {10.1209/epl/i2002-00532-1},
url = {https://doi.org/10.1209/epl/i2002-00532-1},
year = {2002},
month = {jan},
publisher = {},
volume = {57},
number = {1},
pages = {1},
author = {S. Burger and F. S. Cataliotti and C. Fort and P. Maddaloni and F. Minardi and M. Inguscio},
title = {Quasi-2D Bose-Einstein condensation in an optical lattice},
journal = {Europhysics Letters},
}

@article{Hadzibabic2004,
  title = {Interference of an Array of Independent Bose-Einstein Condensates},
  author = {Hadzibabic, Zoran and Stock, Sabine and Battelier, Baptiste and Bretin, Vincent and Dalibard, Jean},
  journal = {Phys. Rev. Lett.},
  volume = {93},
  issue = {18},
  pages = {180403},
  numpages = {4},
  year = {2004},
  month = {Oct},
  publisher = {American Physical Society},
  doi = {10.1103/PhysRevLett.93.180403},
  url = {https://link.aps.org/doi/10.1103/PhysRevLett.93.180403}
}

@article{Andrews1997,
  title = {Propagation of Sound in a Bose-Einstein Condensate},
  author = {Andrews, M. R. and Kurn, D. M. and Miesner, H.-J. and Durfee, D. S. and Townsend, C. G. and Inouye, S. and Ketterle, W.},
  journal = {Phys. Rev. Lett.},
  volume = {79},
  issue = {4},
  pages = {553--556},
  numpages = {0},
  year = {1997},
  month = {Jul},
  publisher = {American Physical Society},
  doi = {10.1103/PhysRevLett.79.553},
  url = {https://link.aps.org/doi/10.1103/PhysRevLett.79.553}
}

@Article{Campbell2012,
  author  = {Ramanathan,Anand and Muniz,Sérgio R. and Wright,Kevin C. and Anderson,Russell P. and Phillips,William D. and Helmerson,Kristian and Campbell,Gretchen K.},
  title   = {Partial-transfer absorption imaging: A versatile technique for optimal imaging of ultracold gases},
  doi     = {10.1063/1.4747163},
  number  = {8},
  pages   = {083119},
  url     = {https://pubs.aip.org/aip/rsi/article/83/8/083119/358919/Partial-transfer-absorption-imaging-A-versatile},
  volume  = {83},
  journal = {Rev. Sci. Instrum.},
  year    = {2012},
}

@article{Tc,
  title = {Bose-Einstein condensation in an external potential},
  author = {Bagnato, Vanderlei and Pritchard, David E. and Kleppner, Daniel},
  journal = {Phys. Rev. A},
  volume = {35},
  issue = {10},
  pages = {4354--4358},
  numpages = {0},
  year = {1987},
  month = {May},
  publisher = {American Physical Society},
  doi = {10.1103/PhysRevA.35.4354},
  url = {https://link.aps.org/doi/10.1103/PhysRevA.35.4354}
}
\end{document}